\documentclass{egpubl}

\JournalPaper 
\electronicVersion

\ifpdf \usepackage[pdftex]{graphicx} \pdfcompresslevel=9
\else \usepackage[dvips]{graphicx} \fi

\PrintedOrElectronic

\usepackage{t1enc,dfadobe}

\usepackage{egweblnk}
\usepackage{cite}
\usepackage{bm}
\usepackage{amsmath}
\usepackage{multicol}
\usepackage{url}

\title[Accelerating Surface Tension Calculation in SPH]%
      {Accelerating Surface Tension Calculation in SPH via\\ Particle Classification \& Monte Carlo Integration}

\author[F. Zorrilla \& J. Sappl \& W. Rauch \& M. Harders]
{\parbox{\textwidth}{\centering F. Zorilla, J. Sappl, W. Rauch, M. Harders 
        }
        \\
{\parbox{\textwidth}{\centering Interactive Graphics and Simulation Group \& Unit of Environmental Engineering\\
         University of Innsbruck, Austria
       }
}
}

\journal{}

\volume{2019}

\EGpagenumber
\newcommand {\nrm} [1]{nrm #1 }
\newcommand {\bf} [1]{\textbf{#1}}

\begin{document}

\teaser{
 \includegraphics[width=0.85\linewidth]{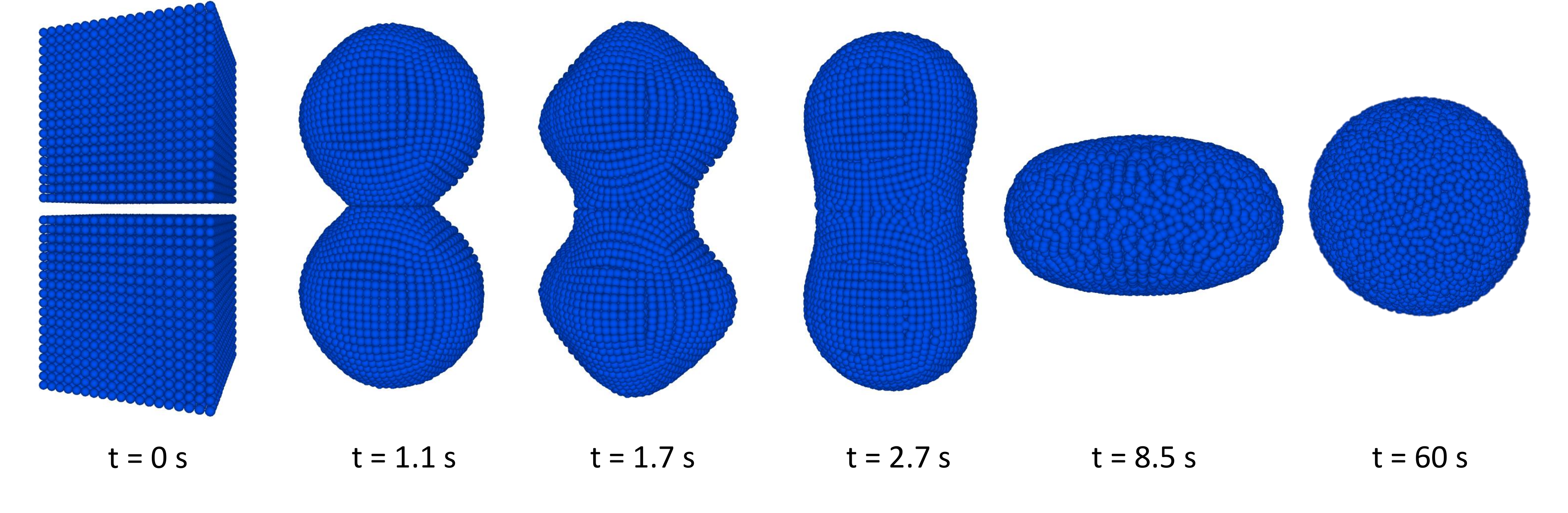}
 \centering
  \caption{SPH time evolution of droplets in 3D (initially cuboid), coalescing into a single spherical droplet. Surface tension calculation is accelerated using our method. Convex as well as concave regions can be robustly handled and shape evolution progresses as physically expected.}
\label{fig:teaser}
}

%
\maketitle

\begin{abstract}
Surface tension has a strong influence on the shape of fluid interfaces. We propose a method to calculate the corresponding forces efficiently. In contrast to several previous approaches, we discriminate to this end between surface and non-surface SPH particles. Our method effectively smooths the fluid interface, minimizing its curvature. We make use of an approach inspired by Monte Carlo integration to estimate local normals as well as curvatures, based on which the force can be calculated. The technique is applicable, but not limited to 2D and 3D simulations, and can be coupled with any common SPH formulation. It outperforms prior approaches with regard to total computation time per time step, while being stable and avoiding artifacts.



\end{abstract}

\section{Introduction}

Surface tension is a phenomenon appearing at the interface of differing media, typically involving a liquid and a gas; such as, for instance, a water-air interface. It results from cohesive forces attracting the molecules of the liquid towards each other. Formally, surface tension is defined as the ratio between the surface force and the distance along which it acts. These forces lead, for instance, to smoothing of fluid surfaces, wherefore they play a vital role in the visual appearance. Accordingly, computational fluid simulations should include estimations of these processes. 

In smoothed particle hydrodynamics (SPH) \cite{monaghan1992smoothed}, fluids are discretized into particles. Due to this, the interface, e.g.~between water and air, is not exactly defined. Therefore, proper ways of approximating the surface tension forces are required. In SPH approaches, these forces are often computed per particle, based on an estimate of the local normal direction as well as of the local curvature. However, some state-of-the-art methods generalize such force calculations to all particles in the fluid, not taking into consideration whether they are located at the surface or not. Technically, this should not introduce any artifacts, since the forces obtained for non-surface particles would be calculated as zero. Nevertheless, computational resources are being spent in the process, without having any effect on the overall fluid behavior.

Instead of the above, it would be advantageous to first classify particles into surface and non-surface ones, as for instance suggested in \cite{he2014robust, sandim2016boundary}. Assuming this can be done efficiently, the subsequent surface tension calculation could be accelerated, leading in total to a reduced computation time per simulation step. Related to this notion, a method for SPH surface detection in 2D has been presented in \cite{dilts2000moving}. They classify a particle as part of the surface, if a circle centered at the particle position is not fully overlapped by circles associated with neighboring particles. Inspired by this idea, we propose an extension of the method, with which we first classify particles (in 2D or 3D). For this step, we employ linear regression, based on machine learning techniques. Once the particles are classified, the local normal and curvature have to be obtained. This is realized by a Monte Carlo approach, where the geometry is locally sampled to determine local coverage. The approach only requires the neighborhood geometry, wherefore it is applicable to any currently existing SPH algorithm for fluid simulation. Furthermore, we also suggest adaptive adjustment of the sample resolution, according to the time step. Comparing our approach to state-of-the-art methods for surface tension force estimation, the total simulation runtimes could be consistently reduced with our approach. As an initial example using our method, see Figure~\ref{fig:teaser} -- the evolution of two particle sets is depicted, arranged initially as two cubes, following our surface tension calculation. Note the coalescence of both parts, including oscillatory movement over time, while also exhibiting concavities. The final equilibrium, minimizing surface tension energy is, as expected, a spherical droplet.


\section{Related Work}

Various approaches to calculate surface tension forces in SPH fluids have been proposed in the past. In earlier work, it was attempted to represent surfaces with a smoothed color field, as seen in \cite{morris2000simulating}, \cite{muller2003particle}, \cite{keiser2005unified} and \cite{kelager2006lagrangian}. The latter is a scalar field, which is initially set to one at particle locations, and zero everywhere else. This permits to obtain estimates of surface normals and curvatures. The latter are calculated as the gradient, as well as the divergence of the gradient of the field, respectively. However, the technique usually leads to a random assignment of normals for particles far from the surface. Moreover, errors in the curvature values result and conservation of the fluid momentum was not ensured. The local nature of our method will reduce problems of normal randomness at locations far from the surface, as well as reduce curvature error.

In \cite{becker2007weakly}, the surface tension force is modeled as a sum of cohesion forces between particles in the same fluid phase. However, the equilibrium of these cohesion forces, as found in simulations, does not always correspond to the correct minimal surface area, as one would expect from a surface tension dominated fluid. The method is also prone to clustering of particles on the fluid surface. To avoid such particle clustering, it was suggested in \cite{tartakovsky2005modeling} to introduce a repulsion force when particles are too close to each other. This was achieved by manually tuning a force profile according to particle separation distance. In our method particle clustering is avoided since we do not use cohesion forces. Related to this, it was stated in \cite{akinci2013versatile} that the surface tension force cannot be estimated as a summation of cohesion forces alone, as observed in nature, since SPH particles represent a fluid on a macroscopic level. Instead, they suggest to combine a cohesion with a surface minimization term. Thus, their force term minimizes fluid surface area, conserves momentum, and prevents clustering. However, forces are manually tuned to attract particles in a certain distance range, while repulsing particles that are too close. In contrast, in \cite{yu2012explicit} the curvature minimization problem is first solved on a mesh that is reconstructed from the SPH particles; and later the results are transferred back to the particles. The authors encountered surface waves that could appear due to a mismatch between mesh vertices and underlying SPH particles. The effect could be reduced in a post-processing step. 

All the methods above treat all particles equally. However, for non-surface particles the resulting force will be zero. Thus, time is spent on calculations that do not have an effect on the simulation. Thus, it may be beneficial to classify particles initially, and then compute forces only for surface particles. One of the first methods that distinguishes between surface and non-surface particles is \cite{he2014robust}. The force is modeled based on the asymmetric neighborhood of particles close to the surface, which leads to asymmetries in the summation of Van der Waals interactions. This yields a force acting on surface particles, proportional to surface curvature.
The work in \cite{sandim2016boundary} introduces a method for surface particle classification based on visual occlusion of particles from different viewpoints.
However, the method is computationally in-tensive and cannot accommodate false positives. In \cite{Zhu:2014:CST:2601097.2601201} surface tension was computed for long, thin objects,

In addition to the above, curvature estimation in general point clouds is also a widely studied topic. In related work, magnitudes proportional to the surface curvature are sometimes computed, but not the exact value itself. This suffers from the similar problem, that also in general point clouds surface curvature may not be exactly defined.
Moreover, most existing work already assumes the availability of a surface-only point cloud, e.g. \cite{foorginejad2014umbrella}, \cite{merigot2009robust}. In contrast, our work starts with particle locations in a volume. 
Finally, also note the relation of the problem to SPH surface reconstruction, e.g via distance fields, such as \cite{Zhu:2005:ASF:1073204.1073298, Yu:2013:RSP:2421636.2421641}.


\section{Methods}

Following the idea of modeling surface tension with a continuum method in \cite{brackbill1992continuum}, we calculate the surface tension force via $
\mathbf{f}^{i}_{st}=-\sigma \kappa^{i}\mathbf{\hat{n}}^{i}$
where $\kappa^{i}$ and $\mathbf{\hat{n}}^{i}$ are surface curvature as well as normal at SPH particle $i$ (note that we employ superscipts to denote particle indices). Further, $\sigma$ is a constant surface tension parameter, measured in $N/m$, that depends on the simulated fluid. As mentioned above, we thus have to approximate curvature as well as normal direction per particle. 

Our proposed method is organized in three major algorithmic steps. First, particles in an SPH simulation are classified into two groups -- surface and non-surface particles. Second, the normal vector is estimated for all surface particles. This makes use of a Monte Carlo technique to locally estimate an integral, taking into account neighboring particles. Due to the probabilistic nature of Monte Carlo computations, the resulting normal vectors are additionally smoothed. Thirdly, following a similar Monte Carlo strategy, we estimate local curvature, again only for the classified surface particles, and again with a subsequent smoothing step. The described process can be applied to 2D or 3D scenes. In addition, the number of random samples, and thus the accuracy, can automatically be adjusted according to the simulation time step. Employing the computed data, the surface tension forces per particle can be computed.  The individual steps are described in detail in the following.

\subsection{Particle classification}

The main idea of classifying particles is to reduce the computation time of the surface tension calculation. We aim to achieve this by excluding (ideally all) non-surface particles from the calculation. Doing so should not affect the overall result since for the latter the surface tension force should be zero anyhow. In contrast, it is crucial for the correctness of the result that no surface particle be misclassified (i.e.,~there should be no \textit{false negatives}). Incorrect classification of non-surface particles as surface ones (i.e.,~\textit{false positives}) should be minimized, but does not affect the correctness of the fluid dynamics.

In order to properly classify the particles, we experimented with defining various feature spaces. Optimally, this should only depend on the local geometry.  As  possible  features,  we  examined, for instance,  the  summation  of  neighborhood  masses,  using  various  weighting  kernels.  However,  it  turned  out  that  good  results could already be achieved by mapping fluid particles into a simple 2-dimensional feature space. The first component of this space is given by the mass-weighted average distance of particles in a local neighborhood: 
\begin{equation}\label{eq5}
\left \|  \frac{\mathbf{X}^{i}}{h}\right \|=\frac{1}{h}\frac{\left \|  \sum_{j}m^{i}\mathbf{x}^{j}-m^{j}\mathbf{x}^{i}\right \|}{\sum_{j}m^{j}},
\end{equation}
where $m$ and $\bf x$ are masses and positions of particles $i$ and $j$, respectively. The neighborhood is defined by the (user-selected) SPH kernel radius $h$; thus each particle $i$ has associated neighbor particles $j$, at distances smaller than $h$. Note that we normalize by dividing the mass-weighted average by $h$, thus making the feature independent of kernel size. For the second feature dimension, we just employ the number of neighbors per particle $N^i$.

\begin{figure}[t!]
	\centering
	\includegraphics[width=\columnwidth]{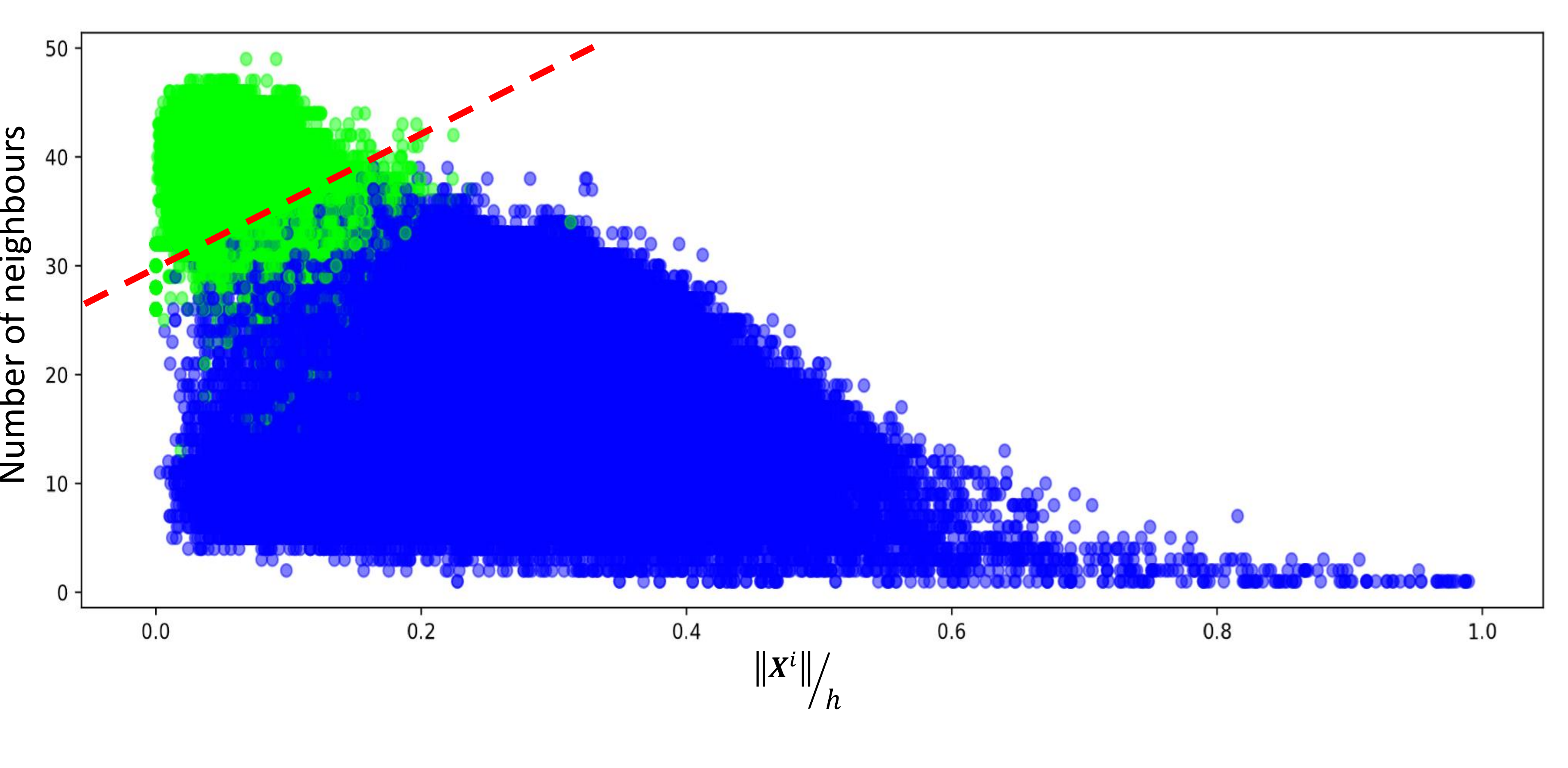}
	\caption{Setup of 2D feature space for particle classification. Each point represents a surface (blue) or non-surface particle (green), for the test simulation data. The dashed red line indicates the linear classifier, which was shifted such as to result in no false negatives. Note that some false positives are still encountered. \label{regression}}
\end{figure}

Next, in order to train a classifier, we have to generate fluid simulation data, and  determine for each particle which class it belongs to. 
The latter training data classification step is done employing a similar strategy as followed for our normal and curvature estimation, as outlined in Sections~\ref{sec:normals} and \ref{sec:curvature}. 
We randomly generate samples on a sphere enclosing a particle and determine coverage of these by neighboring spheres.
In order to achieve high accuracy in this classification, we employ a very large number of samples. 
Further details of the underlying Monte Carlo strategy will be presented below. Simulations to create the training data have been performed using the SPlisHSPlasH framework \cite{SPlisHSPlasH}. 
Scenes with particle numbers between 2K and 30K are employed. 
Different particle configurations, obstacles, boundaries, and gravity forces are used, to ensure broad coverage. Thus generated, and initially classified, particles are plotted in our 2-dimensional feature space in Figure~\ref{regression}. 
Note that using the described features, the two particle classes already exhibit a reasonable separation. It becomes apparent that a linear classifier may already suffice for the classification task. 

For the classification step machine learning strategies can be employed. Since we initially worked in higher-dimensional feature spaces, we decided to employ a neural network classifier. However, as discussed above, moving to a 2-dimensional feature space turned out to be  sufficient for our purpose. We still employ a neural network as linear classifier, however, using a simpler approach, such as for instance support vector machines would also be adequate. In this context, it should be noted that some recent work explored the use of machine learning in fluid simulation, however, only for obtaining solutions to the Navier--Stokes equations, instead of performing the task of classifying particles
(e.g.~\cite{tompson2016accelerating,chu2017data,wiewel2018latent,jeong2015data}). 

In our technique, we effectively obtain a line separating the two classes in the feature space. 
However, since we strive to minimize (i.e.,~optimally avoid) false negatives, we opted to shift the line in normal direction, such that no false positives remain (i.e.,~with respect to the training data). 
The obtained linear classifier is then applied in any new fluid simulation, dividing particles into surface and non-surface ones, progressively per time step. 
Applying this approach in our tests, we did not encounter any false negatives in these simulations, also with different particle configurations and geometries. 
Still, false positives do result. 
In the experiments outlined in Section~\ref{sec:results} the method yielded on average 0.014\,\% false positives in the Droplet, 5.66\,\% in the Dambreak, and 4.75\,\% in the Crown splash test scenario, respectively. Still, the method proved to work fast and be robust with regard to false negatives.
The performance of the method is three orders of magnitudes faster than the timings reported by \cite{sandim2016boundary} for a double dambreak scenario. 
Finally, note that if the classifier is not shifted, then surface particles would be falsely classified as non-surface particles. 
This leads to computational errors, which become visible e.g. as bumps on sphere surfaces; in the end unwanted oscillations and particle motions would result. 
In future work, we will explore the performance of the method also in the context of multi-scale SPH models, i.e. when quite different sampling densities are employed.

\subsection{Normal calculation}
\label{sec:normals}

Once the classification has been finalized, we have to compute the surface normals, as well as the curvatures, per particle. Since we do not make use of any smoothed field in the fluid we have to calculate both values using only the geometry as input. Both calculations follow a similar notion, wherefore, the general idea of both will be outlined first. The following will address the 3D case, but the concept applies analogously to 2D.

The key idea in both cases is to first assume a sphere $S_1$ of radius $r_1$ around a considered surface particle at position ${\bf x}^i$. 
The radius will always be selected equal to the SPH kernel size $h$. 
Next, additional spheres $S_{2}^{j}$ of radius $r_2$ are considered, with their respective centers given by all neighboring particles at position ${\bf x}^j$ (i.e., all surface and non-surface ones combined). 
For this, the neighborhood of a particle is again given according to kernel radius $h$. 
Also, note that $r_2$ is usually smaller than $r_1$. 
The neighboring spheres $S_{2}^{j}$ will overlap the initial sphere $S_1$, located at particle $i$, thus leaving a smaller spherical area $A_1$ that is not overlapped, i.e.,~not within the neighbor spheres.  

Since we work with incompressible or weakly compressible SPH particle distributions the density of the point cloud has to be nearly constant. 
Thus, it can be conjectured that the surface normal ${\bf n}^i$ at the particle will point towards the centroid of the non-overlapped spherical area on the sphere. 
In addition, as will be discussed in more detail below, we also found that the fraction of the sphere that has not been covered is related to the surface curvature at that point.
The area of the sphere that is not overlapped by the neighboring ones can be calculated with a spherical integral. 
However, this integral can be computationally very expensive to determine exactly, wherefore we propose to estimate it using a Monte Carlo integration strategy. 

\begin{figure*}[ht!]
	\centering
	\includegraphics[width=0.85\textwidth]{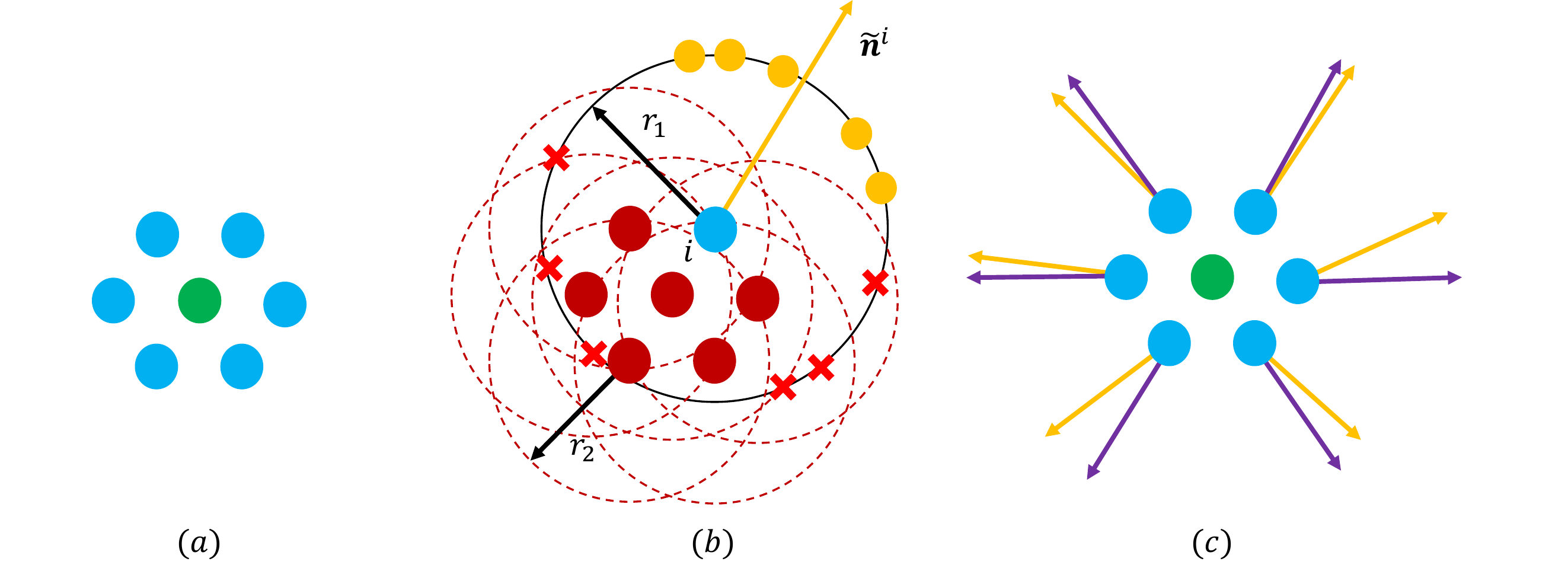}
	\caption{Visualization of normal estimation and smoothing in 2D. $(a)$: point cloud with surface (blue) and non-surface (green) particles. $(b)$: samples ${\bf p}^{k}$ on circle around surface particle $i$; red crosses are overlapped by neighbor circles; yellow dots not, thus these are used for normal estimation. $(c)$: initially estimated normals (yellow) and normals after smoothing (purple)\label{DKVisual}}
\end{figure*}

With regard to the normal computation, we first generate $\mathit{N^i}$ uniformly distributed random sample points $\mathbf{p}^{k}$ on the surface of sphere $S_1$ of particle $i$. Next, we will only consider those that are not inside of any neighboring sphere $S^{j}_{2}$. For our following derivations, we will represent this with a binary function:
\begin{align}\label{eq6}
S(\mathbf{p}^{k})=\begin{cases}
0 & \text{ if } \mathbf{p}^{k}~ is~ overlapped, \\ 
1 & \text{ if } \mathbf{p}^{k}~ is~ not~ overlapped.
\end{cases}
\end{align}
Based on this, we obtain a first estimate of the surface normal:

\begin{equation}\label{eq7}
\tilde{\mathbf{n}}^i=\nrm\left(\sum_{k=1}^{N^i} (\mathbf{p}^{k}-\mathbf{x}^{i})~S(\mathbf{p}^{k})\right), 
\end{equation}
where $\nrm(\cdot)$ is a normalization operator returning a unit vector. Elements in this computation process are visualized in Figure~\ref{DKVisual}$(b)$. Also note that non-surface particles will be assigned with a zero vector. Due to the probabilistic nature of our method, discontinuities in the estimated normal field may be encountered; especially, at lower sampling numbers. However, the normal field should be as smooth as possible on the surface of the fluid.  Therefore, we propose to carry out an additional smoothing step. First, we compute a weighted average of all neighboring surface particle normals, based on the results obtained in the previous step:
\begin{equation}\label{eq21}
\tilde{\textbf{n}}_{Nei}^{i}=\sum_{j=1}^{N^i}W(||{\bf x}^{j}-{\bf x}^{i}||)~\tilde{\textbf{n}}^{j},
\end{equation}
employing a weight kernel $W$, again with kernel radius $h$:
\begin{equation}\label{eq22}
W(x)=\begin{cases}
0 & \text{ if } x>h,\\ 
1-\left | x \right | / h & \text{ if } x\leq h.
\end{cases}
\end{equation}

Also note again that the normals of non-surface particles have been set to zero in the previous step. The final smoothed surface particle normal is then obtained by a weighted average of both computed temporary normals:
\begin{equation}\label{eq20}
\hat{\textbf{n}}^i= \nrm\left((1-\tau)~\tilde{\textbf{n}}^i + \tau ~ \tilde{\textbf{n}}_{Nei}^i\right),
\end{equation}
where $\tau$ is a user selected interpolation parameter. For all our computations we have set it to 0.5. 
The outcome of the normal smoothing process is also illustrated in Figure~\ref{DKVisual}. 
Further note that this smoothing step could potentially be repeated several times.

In order to evaluate the accuracy of the proposed normal estimation process we carried out a comparison between analytically defined and our estimated normals. As error measure we determine $\| \hat{\mathbf{{n}}}^{i} - \mathbf{{n}}^{a}\|$, where $\hat{\mathbf{{n}}}^{i}$ is the estimated normal and $\mathbf{{n}}^{a}$ the analytically correct one. 
The latter were both obtained for the geometry of a 2D circle; a random  2D point cloud is generated by sampling the geometry. Next, due to the random nature of our estimation process, we determine as final error value the average of 100 computations. The results of this study are summarized in Table~\ref{NormalError}. 
We show both the dependency of the average error on the number of samples as well as on the number of smoothing steps. As can be seen, the higher the number of sampling points, the more accurate the approximation becomes. Also, additional smoothing improves the estimates, by filtering out noise incurred by the representation as a point cloud
 
\begin{table}[b!]
\caption{Normal estimation error for 2D circle test case. The average error depends on the number of samples in the Monte-Carlo integration as well as on the number of times the smoothing algorithm is applied.}
\label{NormalError}
\begin{center}
\begin{tabular}{clll}
\hline \\ [-2ex]
        \multicolumn{1}{l}{}  & \multicolumn{3}{c}{\bf{Smoothing Steps}}  \\
        \bf{Samples} & \multicolumn{1}{c}{0} & \multicolumn{1}{c}{1} & \multicolumn{1}{c}{2} \\ [1ex]
                      \hline \\ [-2ex]
  \multicolumn{1}{c}{10}  & 0.290  & 0.171 & 0.131\\
  \multicolumn{1}{c}{20}  & 0.201  & 0.115 & 0.085\\
  \multicolumn{1}{c}{50}  & 0.120  & 0.068 & 0.051\\
  \multicolumn{1}{c}{100} & 0.088  & 0.049 & 0.035\\
  \multicolumn{1}{c}{500} & 0.038  & 0.021 & 0.017\\ [1ex]
  \hline
\end{tabular}
\end{center}
\end{table}


\subsection{Curvature calculation}
\label{sec:curvature}

The surface curvature of a 3D surface is locally given by two values, also known as principal curvatures \cite{goldman2005curvature}. These are defined as the eigenvalues of the shape operator at a point on the surface. By averaging the two we obtain the mean curvature $\kappa$. 
Gaussian andmean curvature estimation fail with point clouds including noise.

\newpage
\noindent We have found that it is possible to establish a direct relation between the mean curvature and the fraction of a sphere that is not covered by neighboring ones,  via the process outlined above. We begin by noting that the fraction of the uncovered surface area of a sphere, using the mapping function \eqref{eq6}, is given by:
\begin{equation}\label{eq9}
\lambda = \frac{1}{4\pi r^{2}}\iint S(\mathbf{x}(\theta, \varphi)) r^{2}\sin(\varphi) \,\text{d}\theta \text{d}\varphi,
\end{equation}
where $\mathbf{x}(\theta, \varphi )$ is a sphere surface location with spherical coordinates $\theta$ and $\varphi$. As before, instead of attempting to compute this value exactly, we will approximate it, for a particle $i$, based on random samples following a Monte Carlo integration strategy:
\begin{equation}\label{eq10}
\lambda \approx \frac{1}{N^i} \sum_{k=1}^{N^i}S(\mathbf{p}^{k}).
\end{equation}
As will be seen later, it is possible to estimate $\kappa$ from $\lambda$, which itself can be determined from the randomly sampled points $\mathbf{p}^{k}$. Note that $\lambda\,\in\,\left[ 0, 1\right]
$. We will first derive the underlying relationship in 2D, and later extend to 3D.

\subsubsection{Relationship in 2D}

\begin{figure}[b!]
\centering
	\includegraphics[width=0.88\columnwidth]{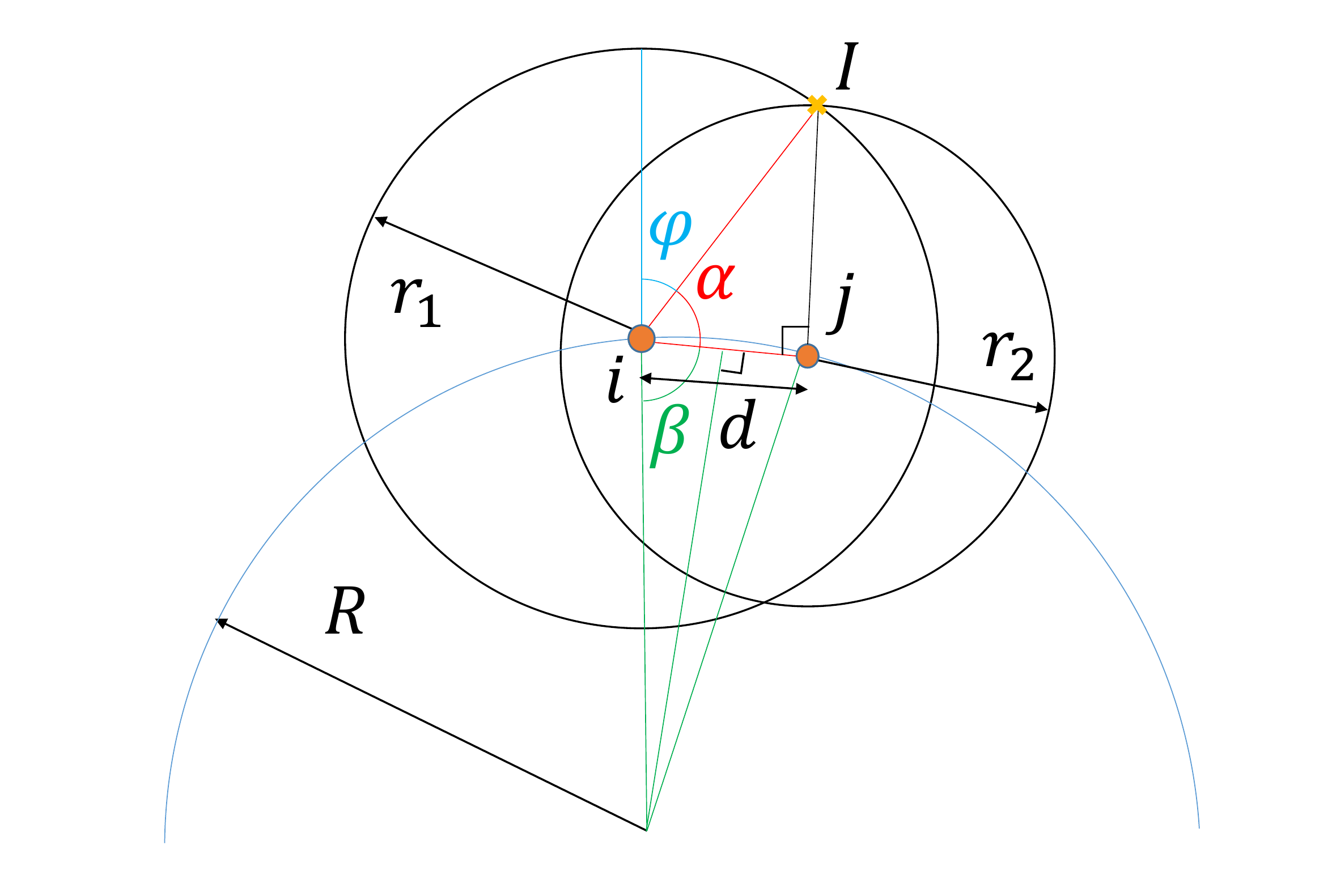}
	\caption{Configuration for maximal arc coverage of circle around neighbor particle j on circle around particle i. \label{Optimal}}
\end{figure}

The following derivation is explained while closely referring to the illustration and notation in Figure~\ref{Optimal}. We start with assuming in 2D a circular outline (shown on the bottom in blue), representing a shape for which the curvature should be determined. The circle has a radius of $\mathit{R}$, and thus the sought curvature $\kappa$ is given in this case analytically by the reciprocal $1/{R}$. However, later the formalism should be applied to any arbitrary shape or curve, based on randomly sampled locations.

First, in order to render our derivation independent of particle size, we will attempt to estimate an adjusted curvature parameter $\tilde{\kappa}=h\kappa$, considering correspondingly also an adapted $\mathit{\tilde{R}}=R/{h}$.
With this in mind, as a starting point for examining in this case the relationship between $\lambda$ and $\tilde{\kappa}$, we will begin with deriving a lower bound $\lambda_{min}$, i.e.,~in 2D the minimal arc that would not be covered by neighboring circles. For this, first consider a particle $i$ on the circular outline. We associate with this particle again a circle $\mathit{C}_{1}$, with radius $\mathit{r}_{1}$, and center ${\bf x}^{i}$. Next, consider additional neighboring particles $j$, akin to what was discussed above; to these again correspond circles $C^{j}_{2}$ with centers ${\bf x}^{j}$, all with the same radius $\mathit{r}_{2}<r_1$, overlapping circle $C_1$. Note that the maximal overlap will result for those particles $j$ that are also located on the circular outline; in 2D there would be two of these, next to particle $i$. Thus, we have to find the geometrical relationship at which circle $C_2$ around such a particle $j$ would cover a maximal arc on $C_1$. 
When the circles overlap, we can find two intersection points; denoting the outer one as ${\bf x}^I$, the maximum coverage will result when the vector between ${\bf x}^I$ and ${\bf x}^j$ is perpendicular to the vector between ${\bf x}^i$ and ${\bf x}^j$ (see also Figure~\ref{Optimal}). In this situation, the angle between the normal at particle $i$ and the vector between ${\bf x}^i$ and ${\bf x}^I$ is given as $\varphi$. Also note that this angle can be obtained via:
\begin{equation}\label{eq11}
\varphi=\begin{cases}
\pi - \alpha -\beta & \text{ if } \tilde{\kappa}>0\\ 
\beta-\alpha & \text{ if } \tilde{\kappa}\leq 0,
\end{cases}
\end{equation}

where angles $\alpha$ and $\beta$ are defined based on the chord between the particle positions, as depicted. Also note that the distance between the latter is given as:
\begin{equation}\label{eq13}
d = \sqrt{r_{1}^{2} - r_{2}^{2}}=h~\sqrt{1 - \left(\frac{r_{2}}{r_{1}}\right )^{2}}.
\end{equation}
According to the geometric configuration, both angles can be obtained via:
\begin{align}\label{eq12.1}
\alpha &= \sin^{-1}\left( r_{2}/ r_{1} \right), \\
\beta = \cos^{-1}\left (  \frac{d/2}{R}\right ) &=\cos^{-1}\left (  \tilde{\kappa}/2\sqrt{1 - \left(r_{2}/r_{1}\right )^{2}} \right ).
\end{align}

Finally, due to having two neighboring particles in symmetric configuration, we have to consider $2\varphi$ for the non-covered arc. Overall, we obtain $\lambda_{min} = 2\varphi / 2\pi$. Using the previous equations, we obtain a closed form solution, independent of the sign of the curvature:
\begin{equation}\label{eq12.3}
\tilde{\kappa}= -2\left(1 - \left( r_{2}/ r_{1} \right )^{2}\right)^{-1/2}\cos\left (\lambda_{min} \pi + \alpha  \right),
\end{equation}
where the adjusted curvature is related to the ratio of the radii $r_2/r_1$ and the minimal covered fraction $\lambda_{min}$, which we approximate via random sampling.


\begin{table*}[t!]
\centering
\begin{tabular}{ccllllll}
\hline \\ [-2ex]
\multicolumn{2}{c}{\bf{Analytical}} & \multicolumn{1}{c}{$0.213$} & \multicolumn{1}{c}{$-0.213$} & \multicolumn{1}{c}{$0.031$} & \multicolumn{1}{c}{$-0.031$} & \multicolumn{1}{c}{$0.5$} & \multicolumn{1}{c}{$-0.5$}\\ [1ex]
\hline\\ [-2ex]
\bf N & \bf{50} & $0.145 \pm 0.365$   &  $-0.273 \pm 0.332$  &  $0.038 \pm 0.325$  &  $-0.049 \pm 0.322$  &  $0.424 \pm 0.275$  &  $-0.634 \pm  0.345$\\
& \bf{200}      & $0.199 \pm 0.125$   &  $-0.241 \pm 0.159$  &  $0.052 \pm 0.142$  &  $-0.014 \pm 0.142$  &  $0.431 \pm 0.173$  &  $-0.584 \pm 0.150$\\
& \bf{1000}     & $0.196 \pm 0.067$   &  $-0.238 \pm 0.070$  &  $0.046 \pm 0.070$  &  $-0.033 \pm 0.068$  &  $0.426 \pm 0.063$  &  $-0.583 \pm 0.085$\\
& \bf{10000}    & $0.194 \pm 0.021$   &  $-0.238 \pm 0.023$  &  $0.030 \pm 0.019$  &  $-0.034 \pm 0.021$  &  $0.428 \pm 0.019$  &  $-0.589 \pm  0.021$\\ [1ex] \hline
\end{tabular}

\vspace{2mm}
\caption{Comparison of analytically defined (top row) with our estimated (bottom four rows) curvatures, obtained for an ellipsoid with semi-axes (a=100, b=200, c=400); the latter is approximated for our method with a random point cloud. Measurements given both for negative(inside) and positive (outside) curvature. Average estimated curvatures and standard deviations are provided, based on 40 measurements,for different numbers of random samples N.}
\label{CurvatureEstimation}
\end{table*}

\subsubsection{Relationship in 3D}

In 3D we follow a similar derivation. As before, we attempt to do this via estimating the ratio of a minimal, uncovered spherical surface area to the complete surface of a sphere. Again, we assume a particle $i$ on this surface, surrounded by several particles $j$, for which again local spheres of radius $r_1$ and $r_2$ are defined. In 3D the uncovered spherical surface area  
$\mathit{A}$ will be a spherical cap, which is given analytically. A cap on a sphere with radius $\mathit{R}$, defined by a projected solid angle $\varphi$ is given as:
\begin{equation}\label{eq15}
A=\int_{0}^{\phi}\int_{0}^{2\pi}R^{2}\sin(\varphi) d\varphi d\theta= 2\pi R^{2}(1-\cos(\varphi)).
\end{equation}

As in the 2D case, we compute $\lambda_{min}$ based on the non-covered surface area:
\begin{equation}\label{eq16}
\lambda_{min}= \frac{2\pi R^{2}(1-\cos(\varphi))}{4\pi R^{2}}= \frac{1-\cos(\varphi)}{2}.
\end{equation}

Thus, rearranging the terms we can also in 3D express the adjusted curvature analytically:
\begin{equation}\label{eq16.2}
\tilde{\kappa}= -2\left(1 - \left(r_{2}/r_{1}\right )^{2}\right)^{-1/2}\cos\left (\cos^{-1} 
\left ( 1-2 \lambda_{min} \right )+ \alpha  \right),
\end{equation}
again depending on the ratio of $r_2/r_1$ and $\lambda_{min}$, which we can estimate. For our implementation and tests we employed the ratio $r_{2}/r_1 = 0.8$, which yielded optimal performance.



\begin{figure}[b!]
	\centering
	\includegraphics[width=42mm]{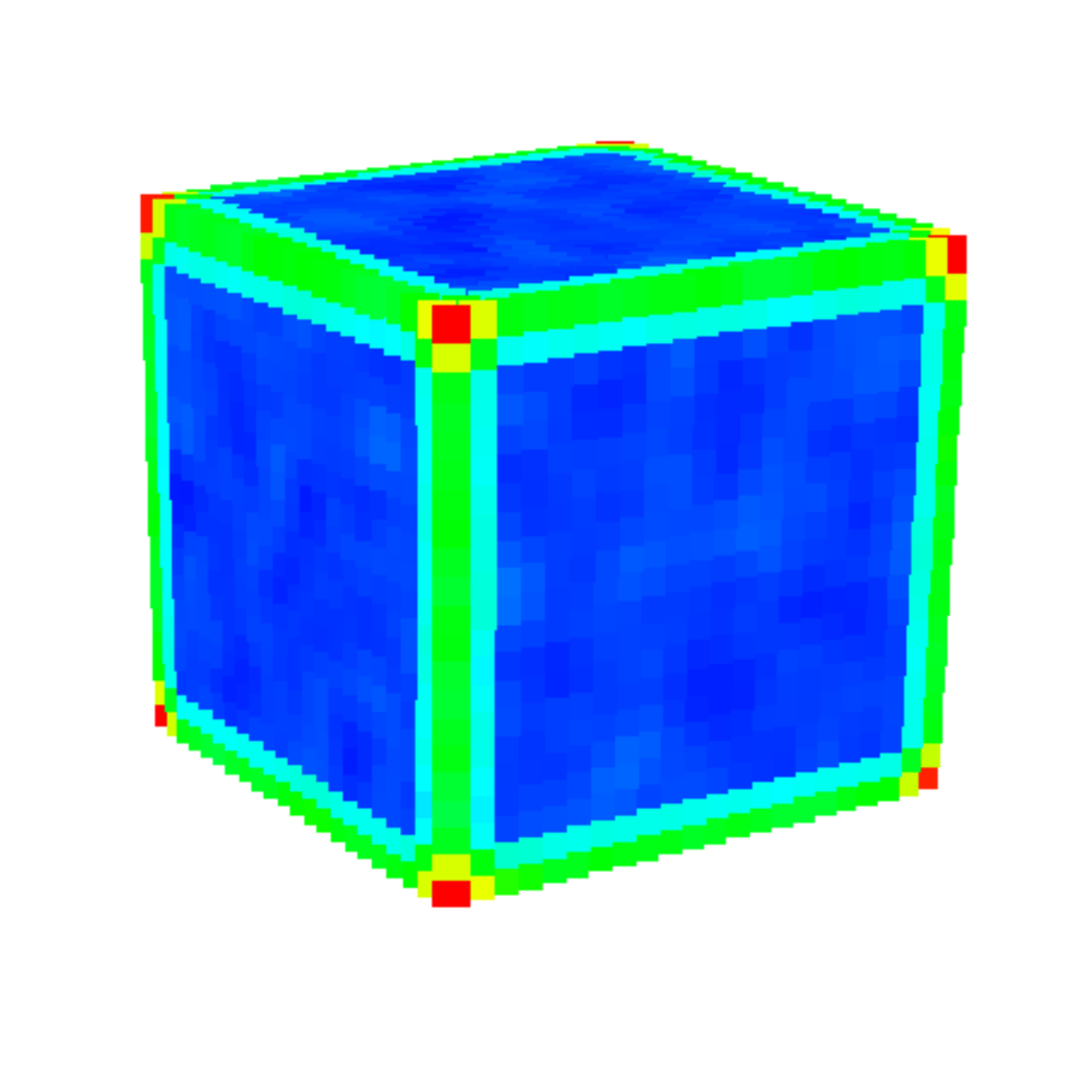}\\
	\includegraphics[width=76mm]{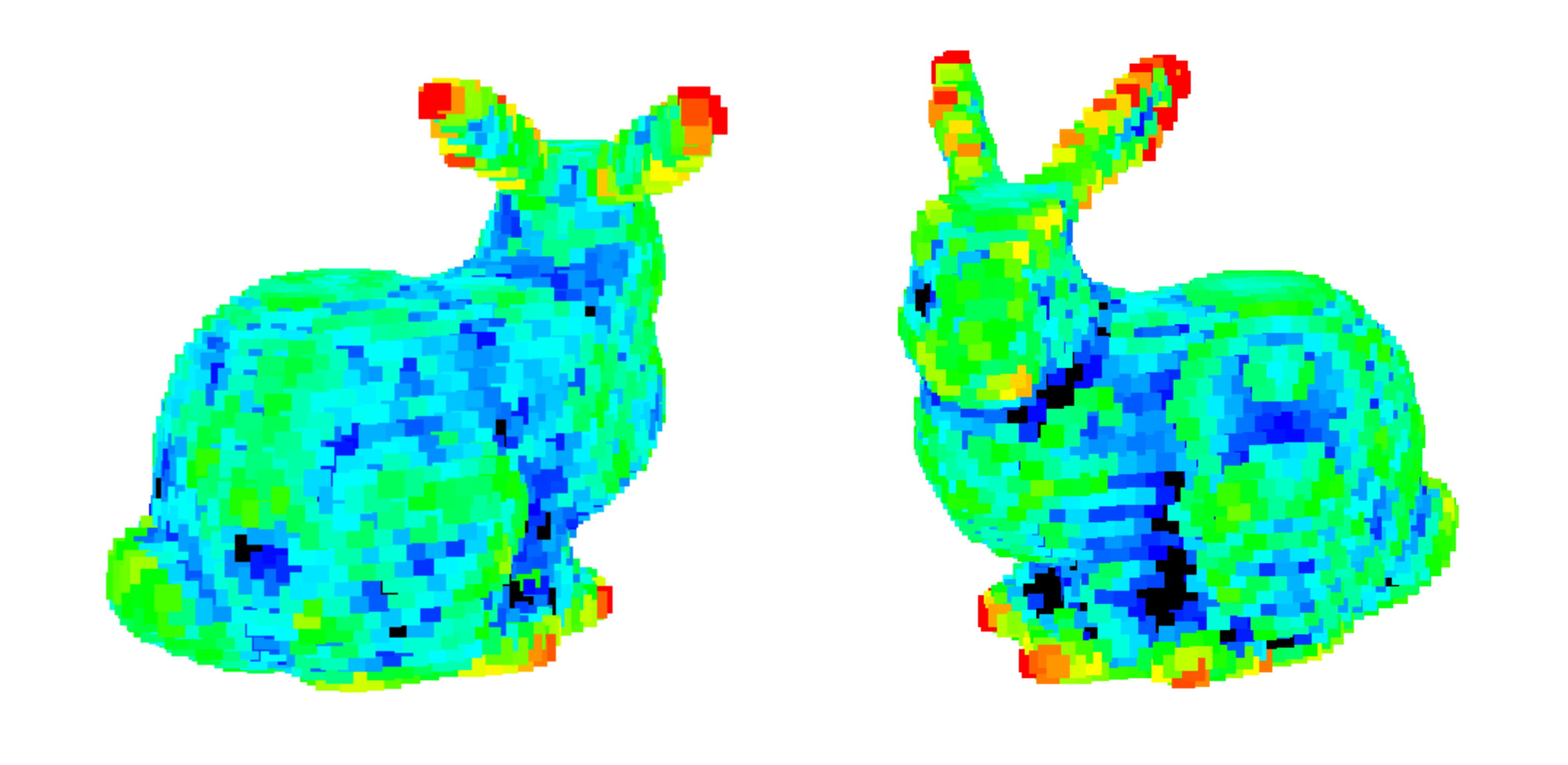}
	\caption{Computed curvature (dark blue = negative) of test point clouds - cube (top); Stanford bunny (bottom). \label{overflow}}
\end{figure}

The described approach can be applied to estimate the surface curvature of any given point cloud. In Figure~\ref{overflow}, we provide examples of curvature calculations, based on our method. Curvatures of a simple cube and a Stanford bunny point cloud are visualized, color-coded from negative (dark blue) to positive (red) values.

In order to evaluate our curvature estimation method, we compare our approximations with analytically defined mean curvature values. The latter can, for instance, be obtained in closed form for any point on an ellipsoid \cite{bektasgeneralized}. 
Thus, we create an ellipsoidal point cloud, for which we obtain our estimate, and compare to the ground truth. 
Due to the stochastic nature of our method, we determine the mean and the standard deviation for 40 measurements. 
Moreover, note that accuracy again depends on the number of samples, wherefore we also tested our method for different amounts of such samples.
The results of this validation are compiled in Table~\ref{CurvatureEstimation}. 
As can be seen, for smaller curvatures our estimation approaches the correct solution, independent of the curvature sign. 
Moreover, even for a small number of samples our estimated average curvature is close to the correct solution. 
Nevertheless, the standard deviation is large for small sample numbers, but can be reduced by increasing the number of samples. 
In addition, we found that for larger curvatures, also larger errors in the mean curvature resulted. 
This is due to the sphere radius $r_1$ becoming closer to actual surface features.

%

\begin{figure*}[h!]
\centering
  \includegraphics[width=0.8\textwidth]{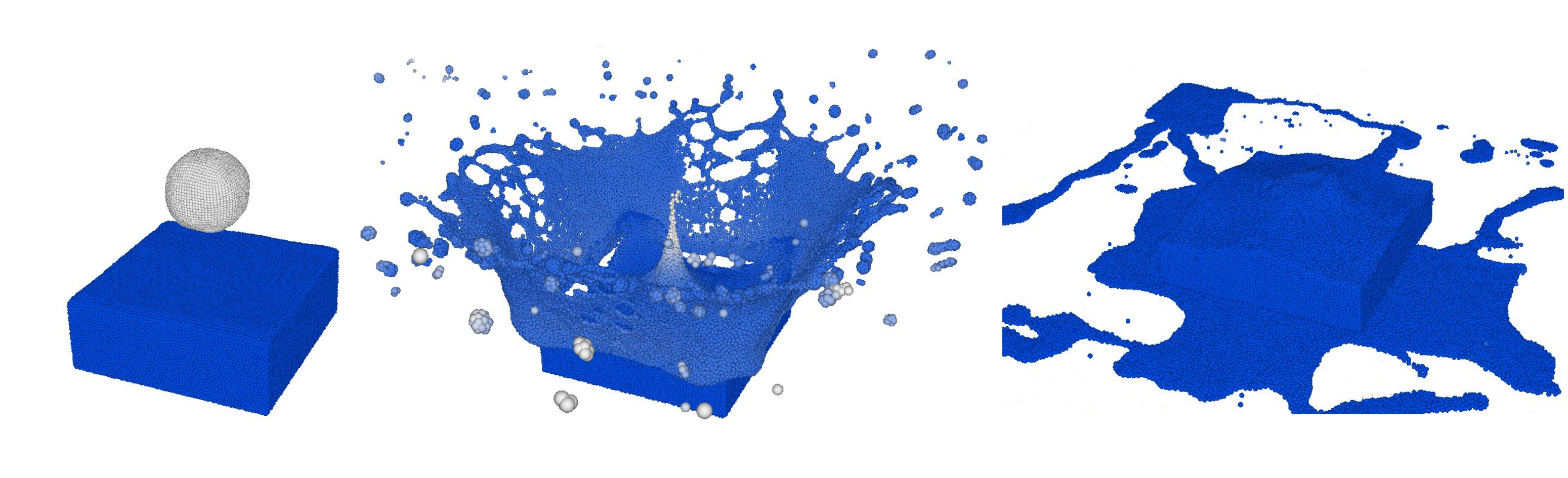}
  \caption{Example of crown splash simulation using DFSPH and our surface tension force estimation (adaptive sampling with $C_{SD}\,{=}\,10,000$, adaptive time step $0.1$--$1\,ms$).}
  \label{Crown}
\end{figure*}

\subsubsection{Curvature smoothing and adaptive sampling}
Similar as for the normal field, the curvature field should also be smooth along the surface. 
The probabilistic nature of the estimation process may also introduce artifacts. Thus, we again suggest to apply one or more smoothing steps, averaging computed curvatures in a local neighborhood.

Furthermore, as already seen in Table~\ref{NormalError}, the accuracy of Monte Carlo approaches will depend on the number of samples. 
A straightforward approach could be to employ a constant number at all times; however, we have found that adapting the number according to the underlying numerical simulation is advantageous. The idea is inspired by the Courant--Friedrichs--Lewy (CFL) condition \cite{monaghan1992smoothed}, which relates numerical time step, spatial discretization, and propagation velocity. 
According to this, solution time steps in SPH algorithms are often adaptively adjusted; commonly based on forces or velocities of the fluid particles. Along this line, we propose to adjust the number of random sampling points used per time step as $N=t_{s}C_{SD}$, with time step $\mathit{t_{s}}$ and user-selected proportionality constant $\mathit{C_{SD}}$. 
The latter can be considered as a sampling density factor, its value representing a trade-off between accuracy and computation speed. 
We have achieved good results with setting this parameter to 10,000--100,000. Our adaptive sampling makes the total number of samples per particle over a simulation time period independent of the numerical time step size.

\begin{table*}[t!]
	\centering 
	
	\begin{tabular}{llrrrrrrr}
	    \hline \\[-2ex]
		\multicolumn{3}{l}{} & \multicolumn{2}{l}{\bf{Becker 2007  \cite{becker2007weakly}}} & \multicolumn{2}{l}{\bf{Akinci 2013\cite{akinci2013versatile}}} & \multicolumn{2}{l}{\bf{Our method}}    \\ \cline{4-9} 
		&  \bf{Scene} &N  &${\bf f}_{st}$ [ms] & Total [ms] & ${\bf f}_{st}$ [ms] & Total [ms] & ${\bf f}_{st}$ [ms] & Total [ms] \\ [0.5ex]
		\hline \\[-2ex]
		DFSPH
		& Droplet   & 10100 & \bf{6.60} &  41.89 &  10.20 &  49.22 &  8.95 &  \bf {38.43} \\
		& Crown     &  145000 & 62.19 & 478.44 & 103.89 & 508.48 & \bf{17.25} & \bf{424.81} \\
		& Dam Break & 26100  & 16.13 & 116.05 &  19.49 & 111.48 & \bf {5.44} &  \bf{74.17} \\ [0.5ex]
		\hline \\[-2ex]       
		PCISPH
		& Droplet   & 10100 &  \bf{7.15} &   98.91 &  11.10 & 129.17 &  9.62 &  \bf{51.09} \\
		& Crown     &  145000 & 87.18 & 1089.16 & 115.36 & 961.35 & \bf{19.61} & \bf{859.49} \\
		& Dam Break & 26100 & 18.01 &  435.23 &  22.11 & 353.40 & \bf{6.86} & \bf{337.70} \\ [0.5ex]
		\hline \\[-2ex]        
		WCSPH
		& Droplet   & 10100 &  7.46 &   26.64 &  11.28 &  30.22 &  \bf{3.96} &  \bf{23.82} \\
		& Crown     &  145000 & 83.93 &  312.13 & 112.73 & 349.34 & \bf{18.28} & \bf{258.49} \\
		& Dam Break & 26100  & 15.67 &   61.90 &  22.38 &  69.13 &  \bf{9.67} &  \bf{56.71} \\ [0.5ex]
		\hline       
	\end{tabular}
	\caption{Average computation times in milliseconds per simulation time step, for surface tension calculation as well as the complete SPH solutions. The lowest values are printed in bold font. Three different SPH solvers were employed (DFSPH, PCISPH and WCSPH), as well as three different test scenes (Droplet, Crown, Dam Break), all computed in SPlisHSPlasH. Our proposed adaptive adjustment of sample numbers is employed; also time steps are adapted dynamically according to the CFL condition.}
	\label{Results}
\end{table*}

\section{Results}
\label{sec:results}

\begin{figure*}[t]
\centering
  \includegraphics[width=0.8\textwidth]{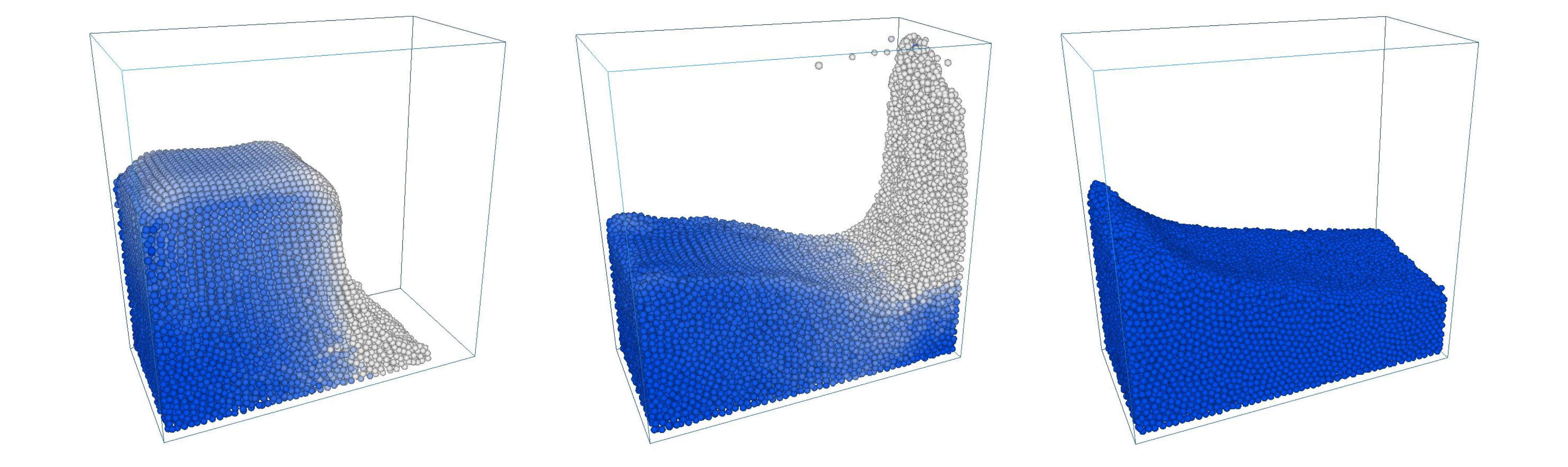}
  \caption{Example of dam break simulation using DFSPH and our surface tension force estimation (adaptive sampling with $C_{SD}=10,000$, adaptive time strp 0.5--2\,ms)}
  \label{Dambreak}
\end{figure*}

In order to evaluate our method we compare it to approaches employed in prior work for the computation of surface tension, specifically the work by Becker and Teschner (2007) \cite{becker2007weakly} and by Akinci, Akinci, and Teschner (2013) \cite{akinci2013versatile}. Further, the surface tension calculations are integrated into different full SPH solvers, specifically weakly compressible SPH (WCSPH) \cite{becker2007weakly}, predictive corrective incompressible SPH (PCISPH) \cite{solenthaler2009predictive}, and divergence free SPH (DFSPH) \cite{bender2017divergence,bender2015divergence}, in a  reference implementation. 
As framework for the comparisons we employ "SPlisHSPlasH" \cite{SPlisHSPlasH}, an open source environment for physically-based simulation of fluids, which provides the implementations for the mentioned comparison algorithms. All computations were performed using only the CPU; i.e.,~no optimizations, such as GPU calculations were employed. Further, we obtain computation times for three different common test scenes, as also suggested by \cite{huber2015evaluation} -- Droplet, Crown splash, and Dambreak. 
These scenes cover various dynamic behaviors, and also require different time step intervals, according to the CFL condition. That is, for Droplet, WCSPH: 
1~ms, DFSPH/PCISPH: 5~ms; Crown splash, all 0.1--1~ms; Dambreak, all 0.5--2~ms. Finally, our proposed adaptive adjustment of sample numbers is employed with $C_{SD}=10,000$. Snapshots of two example simulations of these experiments are shown in Figures~\ref{Crown} and \ref{Dambreak}. 

The timing results of the comparison experiment are provided in Table~\ref{Results}. As can be seen, in all cases our method resulted in reduced average total computation times. Note that in all cases visually highly similar fluid simulation results were obtained, and no instabilities were encountered. 
Moreover, the smaller the time step, the better our method performed compared to the other two surface tension calculation methods, when computing ${\bf f}_{st}$; this becomes especially evident for the Crown splash scene, which employed the smallest time step, where significant improvements resulted for this step. 
However, for larger time steps, the advantage of employing our proposed approach for calculating ${\bf f}_{st}$ is reduced; for instance, in the droplet scene, for both DFSPH and PCISPH, the surface tension calculation times turned out to be slower for our method; nevertheless, the total computation time per time step still remained better. We assume this to be due to non-surface particles also experiencing non-zero surface tension forces in the other methods, which requires additional iterations of the pressure solver to achieve the correct fluid density. 


\section{Discussion \& Conclusion}

We have presented a method to accelerate the calculation of surface tension  forces  in  SPH  fluid  animations.  In  contrast  to  other approaches, we discriminate between surface and non-surface particles. This leads to an improvement in the computation time, since the forces are calculated for just a fraction of the particles. Based on this, we can effectively smooth and minimize the surface of the fluid. The accuracy of our method can be tuned by adjusting the value of $C_{SD}$. When the time step is reduced, e.g. according to the CFL condition, the number of sampling points $N^i$ is also adjusted. Surprisingly, we found that even for a low number, $N^i\,\approx\,10$ the simulation remains stable. It is in cases when the time step is small ($<1\,ms$) that our method offers a considerably improved performance  over  the  other  tested  methods.  However,  when  the  simulation  runs  slower  ($ts\,\approx\,5\,ms$)  the  advantage  is  diminished;  still the computation of our method remains comparable to other algorithms.

A disadvantage we encountered when using a very low resolution in the sampling is the possible creation of incorrect momentum. The sum of the forces around a closed fluid surface should vanish, but for low resolution sampling of the integral this is not ensured. We will examine in future work possible strategies to avoid this artifact. Our approach also includes an estimation of the local mean curvature at the fluid particles; here, the latter could be considered as a general point cloud. Since the surface interface in any point cloud is not clearly defined, the curvature is neither. Related works, 
e.g.~\cite{morris2000simulating,muller2003particle,keiser2005unified}, compute the divergence of the gradient of the color field to estimate the curvature. This leads to a quantity that is only proportional to the exact curvature. In our method we obtain an approximation of the curvature based on a spherical integral. The procedure is similar to searching for a spherical surface, locally best fitting to a point cloud. The mean curvature would be calculated from the radius of such a sphere.

Further, note that instead of employing randomly generated uniform samples, we also explored the use of pseudo-random Halton sequences \cite{berblinger1991monte}. Due to the deterministic nature of the latter, it may be possible to save further computation time. This will be explored in more detail in the future. Finally, our method can effectively be coupled with any other SPH algorithm, since it only takes the geometry as input for the computation. It can be employed to improve the overall SPH computation time, when smoothing fluid surfaces in computer graphics applications. It is left for future work to explore the possibility of applying this type of procedure in other contexts of fluid simulations.

%

\bibliographystyle{eg-alpha-doi}
\bibliography{main.bib}

\end{document}